\pdfoutput=1

\documentclass[11pt]{article}

\usepackage{emnlp2021}

\usepackage{times}
\usepackage{latexsym}
\usepackage{enumitem}
\usepackage{listings}
\usepackage{balance}

\usepackage[T1]{fontenc}
\usepackage[utf8]{inputenc}
\usepackage{multirow}
\usepackage{booktabs} 
\usepackage{microtype}

\usepackage[hang,flushmargin]{footmisc}
\usepackage{microtype}
\usepackage{hyperref}

\newcommand{\ignore}[1]{}

\title{A Few Brief Notes on DeepImpact, COIL, and a Conceptual \\ Framework for Information Retrieval Techniques}

\author{Jimmy Lin \and Xueguang Ma  \\ [1ex]
  David R. Cheriton School of Computer Science \\
  University of Waterloo
}

\begin{document}
\maketitle

\begin{abstract}
Recent developments in representational learning for information retrieval can be organized in a conceptual framework that establishes two pairs of contrasts:\ sparse vs.\ dense representations and unsupervised vs.\ learned representations.
Sparse learned representations can further be decomposed into expansion and term weighting components.
This framework allows us to understand the relationship between recently proposed techniques such as DPR, ANCE, DeepCT, DeepImpact, and COIL, and furthermore, gaps revealed by our analysis point to ``low hanging fruit'' in terms of techniques that have yet to be explored.
We present a novel technique dubbed ``uniCOIL'', a simple extension of COIL that achieves to our knowledge the current state-of-the-art in sparse retrieval on the popular MS MARCO passage ranking dataset.
Our implementation using the Anserini IR toolkit is built on the Lucene search library and thus fully compatible with standard inverted indexes.
\end{abstract}

\section{Introduction}

We present a novel conceptual framework for understanding recent developments in information retrieval that organizes techniques along two dimensions.
The first dimension establishes the contrast between sparse and dense vector representations for queries and documents.\footnote{Consistent with parlance in information retrieval, we use ``document'' throughout this paper in a generic sense to refer to the unit of retrieved text. To be more precise, our experiments are in fact focused on passage retrieval.}
The second dimension establishes the contrast between unsupervised and learned (supervised) representations.
Figure~\ref{table:framework} illustrates our framework.

Recent proposals for dense retrieval, exemplified by DPR~\cite{karpukhin-etal-2020-dense} and ANCE~\cite{Xiong_etal_ICLR2021}, but also encompassing many other techniques~\cite{Gao_etal_ECIR2021_CLEAR,Hofstatter:2010.02666:2020,qu-etal-2021-rocketqa,Hofstatter_etal_SIGIR2021,Lin_etal_2021_RepL4NLP}, can be understood as learned dense representations for retrieval.
This is formulated as a representational learning problem where the task is to learn (transformer-based) encoders that map queries and documents into dense fixed-width vectors (768 dimensions is typical) in which inner products between queries and relevant documents are maximized, based on supervision signals from a large dataset such as the MS MARCO passage ranking test collection~\cite{MS_MARCO_v3}.
See~\citet{Lin_etal_arXiv2020_ptr4tr} for a survey.

\begin{table}[t]
\centering
\begin{small}
\begin{tabular}{l|l|l}
&  {\bf Dense} & {\bf Sparse} \\
\hline
{\bf Supervised} & DPR, ANCE & DeepImpact, COIL \\
{\bf Unsupervised} & LSI, LDA & BM25, tf--idf \\
\end{tabular}
\end{small}
\caption{Our conceptual framework for organizing recent developments in information retrieval.}
\label{table:framework}
\end{table}

Dense retrieval techniques are typically compared against a bag-of-words exact match ranking model such as BM25, which in this context can be understood as unsupervised sparse retrieval.
Although it may be unnatural to describe BM25 in this way, it is technically accurate:\ each document is represented by a sparse vector where each dimension corresponds to a unique term in the vocabulary, and the scoring function assigns a weight to each dimension.
As with dense retrieval, query--document scores are computed via inner products.

What about learned sparse retrieval?
The most prominent recent example of this in the literature is DeepCT~\cite{Dai:1910.10687:2019}, which uses a transformer to learn term weights based on a regression model, with the supervision signal coming from the MS MARCO passage ranking test collection.\footnote{Learning sparse representations is by no means a new idea. The earliest example we are aware of is~\citet{Wilbur_2001}, who attempted to learn global term weights using TREC data, but the idea likely dates back even further.}
DeepCT has an interesting ``quirk'':\ in truth, it only learns the term frequency (tf) component of term weights, but still relies on the remaining parts of the BM25 scoring function via the generation of pseudo-documents.
This approach also has a weakness:\ it only assigns weights to terms that are already present in the document, which limits retrieval to exact match.
This is an important limitation that is addressed by the use of dense representations, which are capable of capturing semantic matches.

These two issues were resolved by the recently proposed DeepImpact model~\cite{Mallia_etal_SIGIR2021}, which also belongs in the family of learned sparse representations.
DeepImpact brought together two key ideas:\
the use of document expansion to identify dimensions in the sparse vector that should have non-zero weights and a term weighting model based on a pairwise loss between relevant and non-relevant texts with respect to a query.
Expansion terms were identified by doc2query--T5~\cite{Nogueira_Lin_docTTTTTquery}, a sequence-to-sequence model for document expansion that predicts queries for which a text would be relevant.
Since the DeepImpact scoring model directly predicts term weights that are then quantized, it would be more accurate to call these weights learned impacts, since query--document scores are simply the sum of weights of document terms that are found in the query.
Calling these impact scores draws an explicit connection to a thread of research in information retrieval dating back two decades~\cite{Anh_etal_SIGIR2001}.

The recently proposed COIL architecture~\cite{gao-etal-2021-coil} presents an interesting case for this conceptual framework.
Where does it belong?
The authors themselves describe COIL as ``a new exact lexical match retrieval architecture armed with deep LM representations''.
COIL produces representations for each document token that are then directly stored in the inverted index, where the term frequency usually goes in an inverted list.
Although COIL is perhaps best described as the intellectual descendant of ColBERT~\cite{Khattab_Zaharia_SIGIR2020}, another way to think about it within our conceptual framework is that instead of assigning {\it scalar} weights to terms in a query, the ``scoring'' model assigns each term a {\it vector} ``weight''.
Query evaluation in COIL involves accumulating inner products instead of scalar weights.

Our conceptual framework highlights a final class of techniques:\ unsupervised dense representations.
While there is little work in this space of late, it does describe techniques such as LSI~\cite{Deerwester_etal_1990,Atreya_Elkan_2010} and LDA~\cite{Wei06}, which have been previously explored.
Thus, all quadrants in our proposed conceptual framework are populated with known examples from the literature.

\section{Comments and Observations}

Based on this framework, we can make a number of interesting observations that highlight obvious next steps in the development of retrieval techniques.
We discuss as follows:

\smallskip
\noindent {\it Choice of bases.}
Retrieval techniques using learned dense representations and learned sparse representations present an interesting contrast.
Nearly all recent proposals take advantage of transformers, so that aspect of the design is not a salient difference.
The critical contrast is the basis of the vector representations:\
In sparse approaches, the basis of the vector space remains fixed to the corpus vocabulary, and thus techniques such as DeepCT, COIL, and DeepImpact can be understood as term weighting models.
In dense approaches, the model is given the freedom to choose a new basis derived from transformer representations.
This change in basis allows the encoder to represent the ``meaning'' of texts in relatively small fixed-width vectors (compared to sparse vectors that may have millions of dimensions).
This leads us to the next important observation:

\smallskip
\noindent {\it Expansions for sparse representation.}
Without some form of expansion, learned sparse representations remain limited to (better) exact matching between queries and documents.
The nature of sparse representations means that it is impractical to consider non-zero weights for {\it all} elements in the vector (i.e., the vocabulary space).
Thus, document expansion serves the critical role of proposing a set of candidate terms that should receive non-zero weights; since the number of candidate terms is small compared to the vocabulary size, the resulting vector remains sparse.
Without expansion, learned sparse representations cannot  address the vocabulary mismatch problem~\cite{Furnas87}, because document terms not present in the query cannot contribute any score.
For DeepImpact, this expansion is performed by doc2query--T5, but in principle we can imagine other methods also.
This leads us to the next important observation:

\begin{table*}[t]
\centering
\begin{small}
\begin{tabular}{lllrl}
\toprule
\multicolumn{3}{l}{\bf Sparse Representations} & {\bf MRR@10} & {\bf Notes} \\
\toprule
& Term Weighting & Expansion \\
\midrule
(1a) & BM25  &  None & 0.184 & copied from~\cite{Nogueira_Lin_docTTTTTquery} \\
(1b) & BM25  & doc2query--T5 & 0.277 & copied from~\cite{Nogueira_Lin_docTTTTTquery} \\
\midrule
(2a) & DeepCT  & None &  0.243 & copied from~\cite{Dai:1910.10687:2019}\\
(2b) & DeepCT  & doc2query--T5 & ? & no publicly reported figure \\
(2c) & DeepImpact  & None &  ? & no publicly reported figure \\
(2d) & DeepImpact  & doc2query--T5 & 0.326 & copied from~\cite{Mallia_etal_SIGIR2021} \\
(2e) & COIL-tok ($d=32$)  & None & 0.341 & copied from~\cite{gao-etal-2021-coil} \\
(2f) & COIL-tok ($d=32$) & doc2query--T5 & 0.361 & our experiment \\
(2g) & uniCOIL  & None & 0.315 & our experiment \\
(2h) & uniCOIL  & doc2query--T5 & 0.352 & our experiment \\[1ex]
\toprule
\multicolumn{3}{l}{\bf Dense Representations} & {\bf MRR@10} & {\bf Notes} \\
\toprule
(3a) & ColBERT & & 0.360 & copied from~\cite{Khattab_Zaharia_SIGIR2020} \\
(3b) & ANCE & & 0.330 & copied from~\cite{Xiong_etal_ICLR2021} \\
(3c) & DistillBERT &  & 0.323 & copied from~\cite{Hofstatter:2010.02666:2020} \\
(3d) & RocketQA & & 0.370 & copied from~\cite{qu-etal-2021-rocketqa} \\
(3e) & TAS-B & & 0.347 & copied from~\cite{Hofstatter_etal_SIGIR2021} \\
(3f) & TCT-ColBERTv2 & & 0.359 & copied from~\cite{Lin_etal_2021_RepL4NLP} \\[1ex]
\toprule
\multicolumn{3}{l}{\bf Dense--Sparse Hybrids} & {\bf MRR@10} & {\bf Notes} \\
\toprule
(4a) & CLEAR & & 0.338 & copied from~\cite{Gao_etal_ECIR2021_CLEAR} \\
(4b) & COIL-full  & & 0.355 & copied from~\cite{gao-etal-2021-coil} \\
(4c) & \multicolumn{2}{l}{TCT-ColBERTv2 + BM25 (1a)} & 0.369 & copied from~\cite{Lin_etal_2021_RepL4NLP} \\
(4d) & \multicolumn{2}{l}{TCT-ColBERTv2 + doc2query--T5 (1b)} & 0.375 & copied from~\cite{Lin_etal_2021_RepL4NLP}  \\
(4e) & \multicolumn{2}{l}{TCT-ColBERTv2 + DeepImpact (2d)} & 0.378 & our experiment \\
(4f) & \multicolumn{2}{l}{TCT-ColBERTv2 + uniCOIL (2h)} & 0.378 & our experiment \\
(4g) & \multicolumn{2}{l}{TCT-ColBERTv2 + COIL (2f)} & 0.382 & our experiment \\
\bottomrule
\end{tabular}
\end{small}

\caption{Results on the development queries of the MS MARCO passage ranking task.}
\label{table:results}
\end{table*}

\smallskip
\noindent {\it Relating DeepCT, DeepImpact, and COIL.}
The upshot of the above analysis is that retrieval techniques based on learned sparse representations should be divided into an expansion model and a term weighting model.
For example, DeepCT performs no expansion and uses a regression-based scoring model.
DeepImpact performs document expansion and uses a pairwise scoring model.
COIL performs no expansion and uses a ``scoring'' model that generates a contextualized ``weight vector'' (instead of a scalar weight).
This breakdown suggests a number of obvious experiments that help us understand the contributions of these components, which we report next.

\section{Experiments}
\label{section:experiments}

Our proposed conceptual framework can be used to organize results from the literature, which are shown in Table~\ref{table:results} on the development queries of the MS MARCO passage ranking task~\cite{MS_MARCO_v3}.
Some of these entries represent figures directly copied from previous papers (with references shown), while others are novel experimental conditions that we report.

The first main block of the table shows retrieval with sparse representations.
Row (1a) shows the BM25 baseline, and row (1b) provides the effectiveness of doc2query--T5 expansion.
In both cases, the term weights are from the BM25 scoring function, and hence unsupervised.
Learned sparse retrieval techniques are shown in row group (2).
Separating the term weighting component from the expansion component allows us to identify gaps in model configurations that would be interesting to explore.
For example, in row (2a), DeepCT proposed a regression-based term weighting model, but performed no expansion.
However, the term weighting model can be applied to expanded documents, as in row (2b); to our knowledge, this configuration has not been publicly reported.

Similarly, DeepImpact combined doc2query--T5 as an expansion model and a term weighting model trained with pairwise loss.
To better understand the contributions of each component, we could run the term weighting model without document expansion, as outlined in row (2c).
This ablation experiment was not reported in~\citet{Mallia_etal_SIGIR2021}, but would be interesting to conduct.

In row (2e) we report the published results of COIL-tok (token dimension $d=32$), which is the sparse component in the full COIL model (which is a dense--sparse hybrid).
Through the lens of our conceptual framework, a number of extensions become immediately obvious.
COIL can be combined with doc2query--T5.
Using source code provided by the authors,\footnote{\url{https://github.com/luyug/COIL}} we trained such a model from scratch, using the same hyperparameters as the authors.
This variant leads to a nearly two-point gain in effectiveness, as shown in row (2f).

In another interesting extension, if we reduce the token dimension of COIL to one, the model degenerates into producing scalar weights, which then becomes directly comparable to DeepCT, row (2a) and the ``no-expansion'' variant of DeepImpact, row (2c).
These comparisons isolate the effects of different term weighting models.
We dub this variant of COIL ``uniCOIL'', on top of which we can also add doc2query--T5, which produces a fair comparison to DeepImpact, row (2d).
The original formulation of COIL, even with a token dimension of one, is not directly amenable to retrieval using inverted indexes because weights can be negative.
To address this issue, we added a ReLU operation on the output term weights of the base COIL model to force the model to generate non-negative weights.
Once again, we retrained the model from scratch using the same hyperparameters provided by the authors.
When encoding the corpus, we quantized these weights into 8 bits to obtain impact scores; query weights are similarly quantized. 
After these modifications, uniCOIL is directly compatible with inverted indexes.
Our experimental results are reported with the Anserini toolkit~\cite{Yang_etal_SIGIR2017,Yang_etal_JDIQ2018}, which is built on Lucene.

It is no surprise that uniCOIL without doc2query--T5, row (2g), is less effective than COIL-tok ($d=32$), row (2e).
However, uniCOIL with doc2query--T5, row (2h), outperforms COIL-tok without needing any specialized retrieval infrastructure---the weights are just impact scores, like in DeepImpact.
These results suggest that contextualized ``weight vectors'' in COIL aren't necessary to achieve good effectiveness---adding expansion appears sufficient to make up for the lost expressivity of weight vectors, as shown in row (2h) vs.\ row (2e).
To our knowledge, our uniCOIL model, row (2h), represents the state of the art in sparse retrieval using learned impact weights, beating DeepImpact by around two points.

The second main block of Table~\ref{table:results} provides a number of comparable dense retrieval results from the literature.
The highest score that we are aware of is RocketQA~\cite{qu-etal-2021-rocketqa}, whose effectiveness beats all known sparse configurations.
Note that  ColBERT~\cite{Khattab_Zaharia_SIGIR2020} uses the more expressive MaxSim operator to compare query and document representations; all other techniques use inner products.

The final block of Table~\ref{table:results} presents the results of dense--sparse hybrids.
\citet{Lin_etal_2021_RepL4NLP} reported the results of dense--sparse hybrids when TCT-ColBERTv2, row (3f), is combined with BM25, row (1a), and doc2query--T5, row (1b).
To this, we added fusion with DeepImpact, uniCOIL, and COIL-tok ($d=32$).
For a fair comparison, we followed the same technique for combining dense and sparse results as \citet{Lin_etal_2021_RepL4NLP}, which is from~\citet{Ma2021ARS}. 
For each query $q$, we used the corresponding dense and sparse techniques to retrieve top-1k documents. 
The final fusion score of each document is calculated by $s_{\small \textrm{dense}} + \alpha \cdot s_{\small \textrm{sparse}}$.
Since the range of the two different scores are quite different, we first normalized the scores into range(0, 1). 
The $\alpha$ was tuned in the range(0, 2) with a simple line search on a subset of the MS MARCO passage training set.

With these hybrid combinations, we are able to achieve, to our knowledge, the highest reported scores on the MS MARCO passage ranking task for single-stage techniques (i.e., no reranking).
Note that, as before, uniCOIL is compatible with standard inverted indexes, unlike COIL-tok, which requires custom infrastructure.

\section{Next Steps}

In most recent work, dense retrieval techniques are compared to BM25 and experiments show that they handily win.
However, this is not a fair comparison, since BM25 is unsupervised, whereas dense retrieval techniques exploit supervised relevance signals from large datasets.
A more appropriate comparison would be between learned {\it dense} vs.\ {\it sparse} representations---and there, no clear winner emerges at present.
However, it seems clear that they are complementary, as hybrid approaches appear to be more effective than either alone.

An important point to make here is that neural networks, particularly transformers, have {\it not} made sparse representations obsolete.
Both dense and sparse learned representations clearly exploit transformers---the trick is that the latter class of techniques then ``projects'' the learned knowledge back into the sparse vocabulary space.
This allows us to reuse decades of innovation in inverted indexes (e.g., integer coding techniques to compress inverted lists) and efficient query evaluation algorithms (e.g., smart skipping to reduce query latency):\ for example, the Lucene index used in our uniCOIL experiments is only 1.3 GB, compared to $\sim$40 GB for COIL-tok, 26 GB for TCT-ColBERTv2, and 154 GB for ColBERT.
We note, however, that with dense retrieval techniques, fixed-width vectors can be approximated with binary hash codes, yielding far more compact representations with sacrificing much effectiveness~\cite{Yamada:2106.00882:2021}.
Once again, no clear winner emerges at present.

The complete design space of modern information retrieval techniques requires proper accounting of the tradeoffs between output quality (effectiveness), time (query latency), and space (index size).
Here, we have only focused on the first aspect.
Learned representations for information retrieval are clearly the future, but the advantages and disadvantages of dense vs.\ sparse approaches along these dimensions are not yet fully understood.
It'll be exciting to see what comes next!


\section{Acknowledgments}

This research was supported in part by the Canada First Research Excellence Fund and the Natural Sciences and Engineering Research Council (NSERC) of Canada.
Computational resources were provided by Compute Ontario and Compute Canada.

\bibliography{anthology}
\bibliographystyle{acl_natbib}

\end{document}